\documentclass{article}

\usepackage{framed,multirow}
\usepackage{amssymb}
\usepackage{latexsym}

\usepackage{url}
\usepackage[table]{xcolor}
\usepackage{longtable}
\usepackage{hyperref}
\usepackage{multirow}
\usepackage{amsmath}
\usepackage{tabularx} % in the preamble
\usepackage[T1]{fontenc}
\usepackage{afterpage}  % load the afterpage package
\usepackage{makecell}
\usepackage{booktabs}
\usepackage{float}% If comment this, figure moves to Page 2
\usepackage{authblk}
\usepackage[authoryear]{natbib}
\usepackage{graphicx}

\begin{document}

\title{Trustworthy clinical AI solutions: a unified review
of uncertainty quantification in deep learning models for medical image analysis}%
\date{}

\author[1,3]{Benjamin Lambert}
\author[2]{Florence Forbes}
\author[3]{Alan Tucholka}
\author[3]{Senan Doyle}
\author[3]{Harmonie Dehaene}
\author[1]{Michel Dojat}

\affil[1]{Univ. Grenoble Alpes, Inserm, U1216, Grenoble Institut des Neurosciences, Grenoble, 38000, France}
\affil[2]{Univ. Grenoble Alpes, Inria, CNRS, Grenoble INP, LJK, Grenoble, 38000, France}
\affil[3]{Pixyl Research and Development Laboratory, Grenoble, 38000, France}

\renewcommand\Authands{ and }

\maketitle

\begin{abstract}
%%%
The full acceptance of Deep Learning (DL) models in the clinical field is rather low with respect to the quantity of high-performing solutions reported in the literature. Particularly, end users are reluctant to rely on the rough
predictions of DL models. Uncertainty quantification methods have been proposed in the literature as a potential response to reduce the rough decision provided by the DL black box and thus increase the interpretability and the acceptability of the result by the final user. In this review, we propose an overview of the existing methods to quantify uncertainty associated to DL predictions. We focus on applications to medical image analysis, which present specific challenges due to the high dimensionality of images and their quality variability, as well as  constraints associated to real-life clinical routine. We then discuss the evaluation protocols to validate the relevance of uncertainty estimates. Finally, we highlight the open challenges of uncertainty quantification in the medical field. 
%%%%
\end{abstract}

%\linenumbers

%% main text
\section{Introduction}
\label{sec:introduction}

These past years, many Deep Learning (DL) medical applications were proposed for the automatic analysis of various imaging modalities, including Magnetic Resonance Imaging (MRI), Computed Tomography (CT), Ultrasound (US) or histopathological images (see \cite{puttagunta2021medical} for a review). To be accepted and routinely used by clinicians, however, these algorithms must provide robust and trustable predictions. This is of particular importance in the context of clinical applications, where the automated prediction may have a direct impact on patient care. Yet, DL models are often considered and used as black-boxes, due to the absence of clear decision rules, as well as to the lack of reliable confidence estimates associated with their predictions \citep{guo2017calibration}. Additionally, DL models proved to be overconfident about their predictions on outliers data \citep{nguyen2015deep}, and very sensitive to adversarial attacks \citep{ma2021understanding}, which suggests a global lack of robustness of this type of models. Due to these limitations, detecting failures or inconsistencies produced by DL models is complex, raising concerns regarding the reliability and safety of using these algorithms in clinical practice \citep{ford2016privacy}. To tackle this essential aspect, several research directions have emerged in order to mitigate the "black-box issue", including Explainable Artificial Intelligence (XAI) and Uncertainty Quantification (UQ). XAI methods \citep{arrieta2020explainable} propose to explain the prediction of the DL model in a way that is understandable to humans. In the context of medical image analysis, an example of XAI approach is the computing of saliency maps showing the image's relevant features identified by the DL model, or example-based explanations consisting in the presentation of cases similar to the one considered, {\it e.g.} medical images of patients with the same condition, \citep{van2022explainable}. However, concerns have been raised concerning the fidelity and intelligibility of the explanations provided by XAI methods, which may give the misleading impression of a better understanding of the black-box \cite{adebayo2018sanity,rudin2019stop}. On the other side, UQ methods \citep{abdar2021review} were developed to quantify the predictive uncertainty of a given DL model. Enhancing an automated prediction with an estimation of its confidence has numerous benefits. First, it allows the identification of uncertain samples that need human reviewing. In a medical setting, this is particularly crucial to prevent silent errors, that may lead to inaccurate diagnosis or treatment. Second, it enables the identification of the model's pitfalls. For example, unconfident predictions can indicate an incomplete training dataset. It gives insights regarding the knowledge captured  by the model, and can be used to extend the training set with supplementary data, if needed. High uncertainty can also reveal anomalies within the input data, which is critical for Quality Control (QC). Overall, UQ increases trust in the algorithm, and facilitates the interaction between the algorithm and the user. Moreover, UQ benefits from strong theoretical foundations and has emerged, from the clinical point of view, as one of the expected property of a deployed AI algorithm  \citep{tonekaboni2019clinicians}. As a result, the medical-imaging community is becoming increasingly interested in incorporating UQ to image processing pipelines in order to highlight model failures or weaknesses. In this work, we propose a comprehensive overview of such an UQ integration in medical image processing pipelines.

\subsection{Research Outline}
Several review articles focusing on uncertainty in DL can be found in the literature. In \cite{abdar2021review}, authors propose a complete review of UQ methods, as well as their various concrete applications. \cite{hullermeier2021aleatoric} focus their article on the definition of the two main categories of uncertainty, namely aleatoric and epistemic uncertainties, in the context of machine learning applications. In \cite{gawlikowski2021survey}, insights about the various sources of uncertainty are presented. Reviews focusing on Bayesian DL \citep{jospin2022hands,wang2020survey} and prediction intervals \cite{kabir2018neural} have been also published. More recently, \cite{zhou2021survey} present a review of the latest advances considering epistemic uncertainty quantification in DL from the perspective of generalization error. While these various works propose a complete overview of UQ methods in DL from a general point of view, we have noticed the lack of reviews focusing on medical image processing applications, where being able to correctly identify the confidence of the model is crucial. \cite{kurz2022uncertainty} presented a first work in this direction, using a corpus of 22 papers. Their study, however, is restricted to medical image classification. With the present review, we propose to extend the latter by presenting a complete review of 130 peer-reviewed papers implementing UQ applications in supervised DL-based pipelines, for both medical image classification and segmentation. We also aim at providing an in-depth discussion of UQ methods' evaluation procedures, as well as pointing out the challenges of the field and potential future directions. Our review differentiates from other previously published ones by the following contributions: 

\begin{itemize}
    \item A review of UQ methods dedicated to DL medical image processing classification and segmentation.
    \item A focus on the proposed metrics for uncertainty estimates evaluation. 
    \item Discussion on the current challenges and limitations of UQ for medical image analysis, and suggestion of future work directions. 
\end{itemize}

\subsection{Organization of this Review}
This report is divided into four sections. Section \ref{sec:preli} introduces the key concepts addressed in this study, namely the application of DL models to medical image classification and segmentation (subsection \ref{dl_in_medimage}), as well as the main notions of UQ  (subsection \ref{language_uq}). Section \ref{sec:methods} presents the most popular UQ methods applied in the context of medical image analysis. Section \ref{sec:eval} then focuses on the evaluation procedures that can be implemented to assess the usefulness of uncertainty estimates. Finally, Section \ref{sec:discussion} proposes a discussion of the current challenges and gaps in the literature in the field of UQ for DL medical image processing.

\section{Framework} \label{sec:preli}

\subsection{Problem setting}
In this work, we focus on supervised learning approaches. With this classical setting, the goal of the DL algorithm is to learn a task $T$ based on a training dataset composed of pairs of input images $x$, and their associated ground truth $y$. This target represents a class in the context of classification (\textit{e.g.}, \emph{healthy}, \emph{pathological}), whereas it consists in a mask for segmentation tasks (\textit{e.g.}, the manual delineation of tumors). By observing multiple examples of pairs of images and their corresponding labels during training, the learning agent estimates the mapping function $p(y|x)$ from the data. 

\subsection{Deep Learning for medical image analysis}\label{dl_in_medimage}
The common approach for supervised DL medical image processing is the training of a Convolutional Neural Network (CNN) using an annotated dataset ({\it i.e.} the ground truth). The building block of CNNs is the convolutional layer, which convolves the input data with learnable weighted kernels. This enables the extraction of features within the image, while being insensitive to the position, scale and shape.

For medical image classification, popular convolutional architectures comprises Residual and Dense CNNs \citep{huang2017densely} or EfficientNets \citep{tan2019efficientnet}. These architectures consist of a succession of convolutional layers that extract features from the image at different scales while reducing its size, thus its spatial resolution. For medical image segmentation, popular choices include U-Net \citep{ronneberger2015u} and its variants, such as Residual U-Net \citep{kerfoot2018left}, V-Net \citep{milletari2016v}, Attention U-Net \citep{oktay2018attention} or Dynamic U-Net \citep{isensee2018nnu}. These segmentation models are composed of two branches, an encoder and a decoder, forming the U shape. The encoder compresses the dimension of the input image, while the decoder decompresses the signal until it recovers its original size. Between the two modules, skip connections are usually added so that the features learned in the encoder part can be used to generate the segmentation in the decoder part. Similarly to medical images that can be either 2-dimensional (e.g. 2D CT, Optical coherence tomography (OCT), microscopy or colonoscopy) or 3D (e.g MRI, 3D CT, PET...), the CNNs can be implemented in 2D or 3D.

During the supervised training stage, the CNN uses images from the training set to produces predictions, which are compared to the ground truth targets in order to estimate the error of the model. To do so, a loss function is introduced to estimate the discrepancy between predicted and true labels. Standard choices for both image classification and segmentation include the cross-entropy loss or focal loss \citep{lin2017focal}. For segmentation tasks, specific loss functions can also be used such as the popular Dice loss \citep{milletari2016v} and variants (Generalized Dice loss \citep{fidon2017generalised} or Tversky loss \citep{salehi2017tversky}). 

In the context of medical image classification, CNNs provide a categorical probability distribution over the different observable classes, by applying a softmax function on the model's output. The final assigned class corresponds to the one having the highest probability. The same process is applied for medical image segmentation, except that the CNN predicts one class per pixel or voxel. UQ aims at completing these predictions with uncertainty estimates, allowing a better interpretation of the results with respect to the model's confidence. In the following section, the main concepts of uncertainty are introduced. 

\subsection{The specific language of uncertainty}\label{language_uq}
Predictive uncertainty, meaning the uncertainty associated with the prediction of a DL model, is typically divided in two parts: model (or epistemic) and data (or aleatoric) uncertainty. 

\textit{Epistemic uncertainty} describes uncertainty arising from the lack of knowledge about the perfect predictor, considering the current input \citep{hullermeier2021aleatoric}. In complex scenarios, there is often not a single model, but rather a multitude of models that can explain the observed data \citep{gal2016uncertainty}. Thus, uncertainty arises regarding the choice of the model parameters. Epistemic uncertainty is considered to be reducible, meaning that it can be reduced by using additional data. In practice, epistemic uncertainty is expected to be high for images far from the training data distribution (referred to as out-of-distribution (OOD) samples). Such discrepancy between test and training datasets is frequent in medical image analysis, where there may be significant variations between images acquired at different hospitals or using different machines. Additionally, unexpected patterns can be encountered in test images, such as diseases not encountered during training, or artifacts. Popular approaches to improve the generalizability of models to unseen domains include data augmentation \citep{DAchen2020realistic, DAouyang2021causality, DAzhang2020generalizing} or transfer learning \citep{ghafoorian2017transfer}.

\textit{Aleatoric uncertainty} describes intrinsic noise and random effects within the data \citep{hullermeier2021aleatoric}. It is not intrinsic to the model, but rather a property of the underlying generative distribution of the data. In the context of classification or segmentation, aleatoric uncertainty increases when the number of classes is high and when these classes are fine-grained \citep{malinin2019uncertainty}. Aleatoric uncertainty is considered to be irreducible, meaning that it cannot be reduced with more data. Actually, the only way to diminish aleatoric uncertainty would be to increase the measurement system precision to reduce noise that corrupts the dataset \citep{gal2016uncertainty}. Finally, aleatoric uncertainty can be further split into two categories: \textit{homoscedastic uncertainty}, which is identical for each sample of the dataset, and \textit{heteroscedastic uncertainty}, which depends on the query input. 

Lastly, closely linked to this notion of data uncertainty, the notion of \textit{label uncertainty} was introduced for segmentation tasks. It has been observed that inter-rater variability in the context of manual delineations of medical images was important \citep{becker2019variability,joskowicz2019inter}. This has a direct impact on the model's overall uncertainty as the same object of interest (\textit{e.g.} a brain tumor) may have significantly different ground truth delineations depending on the rater. 

In the next section, we propose an overview of the most employed UQ methods for medical image classification and segmentation, in light of the selected corpus of papers. 

\section{Review of Uncertainty-Quantification methods for medical image analysis using Deep Learning} \label{sec:methods}

\begin{figure*}[ht!]
\includegraphics[width=\textwidth]{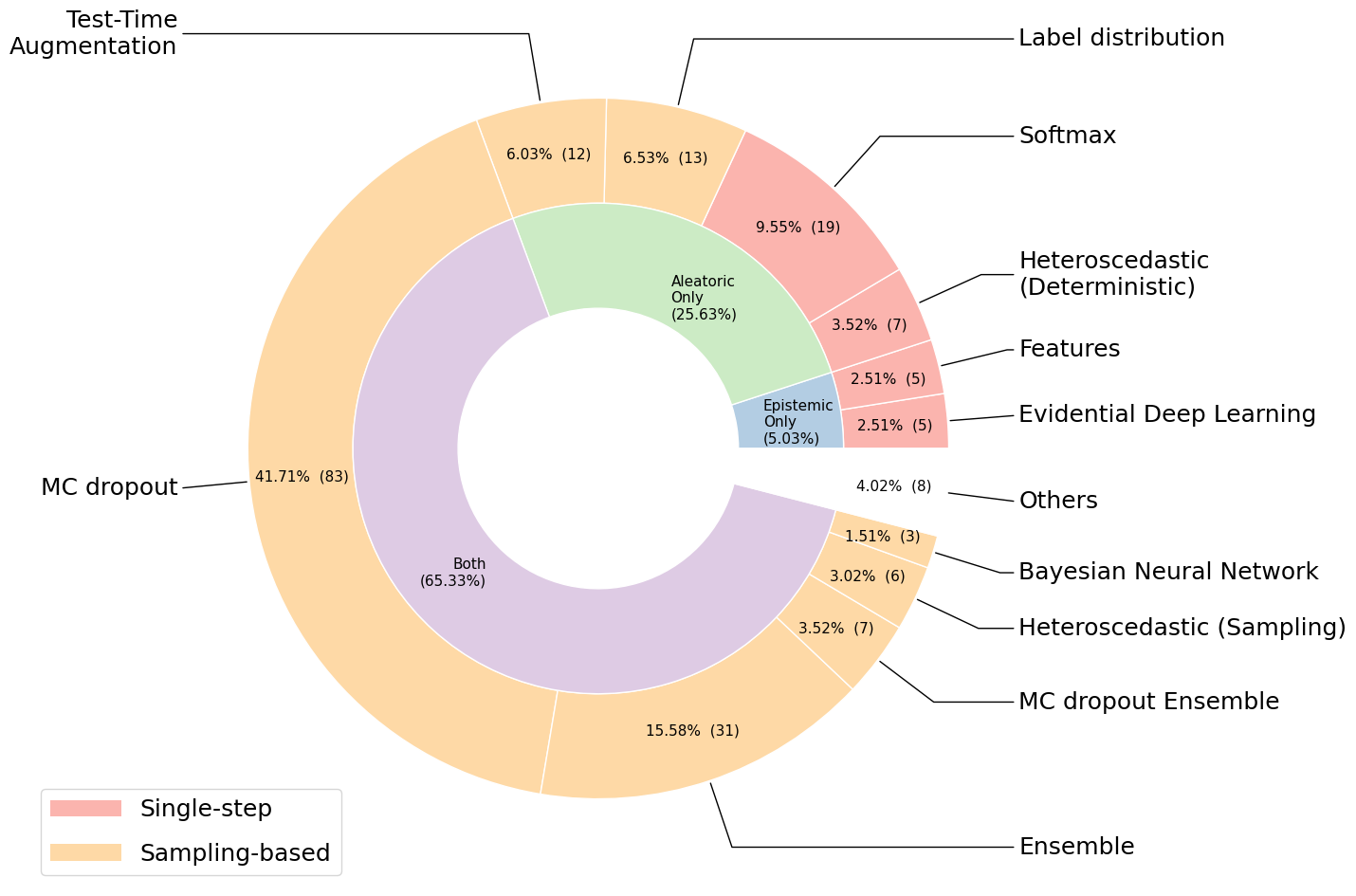}
\caption[]{Implemented UQ methods in the 130 selected papers. The percentage (and the number)  of the selected papers for each class of methods is indicated in the outer ring and the corresponding percentage below the class name. Orange identifies sampling-based methods, and red single-step methods. The inner ring classifies methods according to the type of uncertainty modeled: aleatoric, epistemic or both.} 
\label{fig:method_pie}
\end{figure*}

% Proposition
We performed a systematic search on June 2022 using Google Scholar and PubMed to identify DL studies implementing UQ methods for medical image classification and segmentation published from 2015 (included) to June 2022. The following combination of keywords was used for the search: "Deep Learning", "Uncertainty", "MRI", "CT", "PET", "X-RAY", "Medical image". Studies were included if they 1) implemented supervised DL models for medical image classification or segmentation; and 2) proposed a quantification of the uncertainty of their algorithms. The following exclusion criteria were applied 1) non-peer-reviewed studies (exception were made for papers with more than 30 citations); 2) non-English papers; 3) review articles and 4) animal studies. 130 papers were finally selected for analysis. It resulted a total of 199 UQ models, implemented either as principal contributions or as comparison methods (the exhaustive list of methods can be found in \ref{table:methods}). We first clustered them according to the method used for uncertainty estimation. We further proposed a categorization of these methods according to the type of uncertainty that is modeled, namely epistemic or aleatoric. Moreover, for real-world deployment of a DL model in a clinical setting, speed is crucial for integration into the routine. This means that the implementation of a UQ protocol should not come with prohibitive inference time or computational cost. Thus, we further distinguish between sampling methods, which require multiple inferences per input image (and thus tend to be slow and/or computationally costly), and single-step methods, which produce uncertainty estimate at the cost of a single inference step (and thus tend to be faster). The resulting taxonomy is presented in Figure \ref{fig:method_pie}. In the following of the section, we briefly present each UQ framework. 

\subsection{Softmax methods}\label{section:softmax_methods}
An immediate and simple approach to obtain uncertainty estimates from classification or segmentation by neural network (NN) techniques is to consider the output predicted probabilities $p(y|x)$. Naively, the higher this probability, the more certain is the prediction. Works have been proposed to improve the calibration of the probabilities predicted by a DL model, and ensure that these scores match the true performance of the model  \citep{guo2017calibration,kumar2019verified,nixon2019measuring}. More formally, a model is calibrated if,  for each prediction with the associated probability p, the model is correct 100 x p of the time. However, UQ based on softmax probabilities only consider the distribution over the model's outputs, and not on the model's weights. Thus, this type of deterministic uncertainty estimates only consider aleatoric uncertainty \cite{hullermeier2021aleatoric, kendall2017uncertainties}. 

\subsection{Bayesian Neural Network methods}
In Bayesian Deep Learning (BDL), each weight of the NN is replaced by a distribution, rather than having a single fixed value \citep{blundell2015weight}. To achieve this, a prior distribution $p(w)$ (usually Gaussian) is first initialized over the NN weights. It follows that each weight is represented by a mean and a variance (thus doubling the number of parameters of the model). Then, during training, the model learns the posterior distribution $p(w \vert D)$ given the training dataset $D$ and the prior distribution, which account for the less and more likely parameters given the observed data. The trained Bayesian Neural Network (BNN) is akin to a virtually infinite ensemble of NNs, where each instance has its weights drawn from the learned posterior distribution. During inference, the distribution is marginalized by repeatedly sampling weights from the shared distribution and averaging the predictions. Uncertainty estimates such as the entropy of the predictive distribution, its variance or its mutual information can be computed. BNN places a distribution on the model's outputs as well as on the model's weights, hence is able to model both aleatoric and epistemic uncertainties. While being theoretically founded, BNN requires extensive changes to both the model architecture and training paradigm, and also significantly increases the computational cost of training and inference. 

\subsection{Monte Carlo dropout methods}
In \cite{gal2016dropout}, authors demonstrated that a NN trained with dropout is able to efficiently approximate Bayesian inference without the associated prohibitive computational cost. Based on this principle, the Monte Carlo Dropout (MC dropout) technique proposes to train a model with dropout and keeps it activated during inference. For a given query input, multiple forward passes are then performed. Each time, a different dropout mask is randomly sampled, producing different predictions. Following this process, a predictive distribution is obtained, similarly to BNN. MC dropout allowing to approximate a BNN in any network trained with dropout, it thus rapidly gained popularity. As a counterpart, finding the optimal dropout strategy (rate and position within the NN) is not straight-forward \citep{jungo2020analyzing}.

\subsection{Ensemble methods}
Deep Ensemble (DE) \citep{lakshminarayanan2017simple} proposes to sequentially train a series of NN. As the weights of the neural network are initialized randomly, the models reach different optimums during training. As a result, they produce diverse predictions for the same query input. As for BDL and MC dropout, uncertainty estimates can then be extracted from the ensemble's predictive distribution. A DE does not require any changes to model architecture or training paradigm. Yet, it requires to repeat the training several times, and the aggregation of each individual prediction at inference, which increases the computational cost of this approach. Finally, it is worth noticing that some works propose to associate ensemble and MC dropout in order to get the best of both methods. This allows the combination of individual models uncertainty (though Monte Carlo dropout) as well as the overall ensemble uncertainty (though Deep Ensemble). 

\subsection{Heteroscedastic model-based methods}
As its denomination indicates, an heteroscedastic model aims at evaluating the heteroscedastic part of aleatoric uncertainty. Within this framework, uncertainty is directly learned during training from the data itself, without the need for ground truth labels for uncertainty. Heteroscedastic models can be categorized into two subtypes: sampling methods, which extend MC dropout models, and deterministic approaches, which are a recent improvement and produce the prediction together with its related uncertainty in a single-step. 

\subsubsection{Sampling Heteroscedastic models}
In \cite{kendall2017uncertainties}, authors hypothesize that the network output logits are corrupted by Gaussian noise with mean equal to 0 and variance $z$. The higher the variance, the higher is the aleatoric uncertainty. A model can then be trained to predict the mean logits $\rho$, as well as the noise variance $z$. To do so, authors duplicate the outputs of a MC dropout network: one for the logits, and one for the variance. At each training step, the loss of the model is evaluated by integrating over multiple samples of noise. Paired with the MC dropout framework, this sampling formulation of heteroscedastic models enables the modeling of both epistemic and aleatoric uncertainties.

\subsubsection{Deterministic Heteroscedastic models}
Recently, deterministic variants of the heteroscedastic model have been proposed. In this approach, uncertainty is still learned during training from the data itself, but do not require the integration of the loss over multiple samples of noise. As in the sampling approach, the NN is modified by adding a dedicated output for uncertainty. Then, an uncertainty-augmented loss function is used to learn the predictive task (classification or segmentation), while also learning to predict high uncertainty scores for samples that are likely to be incorrect. In this context, aleatoric uncertainty is learned without any additional cost, as it simply exploits the ground truth label (class or segmentation). 

\subsection{Label-distribution model-based methods}
A branch of the UQ literature focuses on modeling label uncertainty in the context of image segmentation. These approaches focus on datasets for which multiple expert manual segmentations are provided for each image, interpreting the inter-rater variability as a form of ground truth uncertainty. In this setting, it becomes possible to approximate the expert label distribution using generative segmentation neural networks \citep{kohl2018probabilistic}. At inference, sampling from the learned distribution produces diverse segmentation masks, which reproduce the inter-rater variability. We refer to this family of methods as label-distribution models. Despite being intuitive, it is not clear whether or not inter-rater variability can be used as ground truth for uncertainty. In the context of medical image segmentation, there are many cases where a unique segmentation cannot be obtained, for instance due to partial volume effect observed in MRI at the boundaries between healthy tissue and lesions. In that context, experts segmentations exhibit somewhat random variations around the boundaries of the target object. Moreover, experts can over-segment or alternatively under-segment the same object of interest, based on their annotation style. This inter-rater variability is thus rather linked to contextual biases (e.g, radiologist experience or annotation habits) rather that on the true uncertainty of the label \citep{mehta2022qu}. 

\subsection{Test-Time Augmentation}
Test-Time Augmentation (TTA) \citep{ayhan2018test} was proposed as an UQ method to evaluate aleatoric uncertainty. At test time, multiple variants of the input image are generated using Data Augmentation. This can include spatial transformations (\textit{e.g.} flipping, rotation) as well as intensity augmentations (\textit{e.g.} contrast modification, noise injection, or artifacts). This process aims at exploring the impact of input-image transformations on the prediction. Using TTA, the model generates a set of predictions for the same initial input image. From this distribution of predictions, uncertainty metrics can be extracted such as the mediane or the variance. 

\subsection{Feature-based methods}
From a practical point-of-view, epistemic uncertainty is expected to be high for Out-of-distribution (OOD) images, \textit{e.g.} images that are far from the training image distribution. Based on this concrete application, efficient epistemic-uncertainty techniques were recently proposed to detect OOD from the feature map signature of a trained NN \citep{postels2021practicality}. This builds on the hypothesis that feature maps contain information regarding the correctness of a prediction. Despite being efficient for OOD detection, it has been observed that feature-based uncertainty estimates are generally poorly calibrated \citep{postels2021practicality}. These methods are computationally efficient, however their application to medical images remains rare. 

\subsection{Evidential Deep Learning}
The Dempster–Shafer Theory of Evidence (DST) is a framework for dealing with epistemic uncertainty \citep{dempster1968generalization}. In a K-class classification (respectively, segmentation) problem, DST proposes to assign belief masses to each possible class, as well as an overall uncertainty mass. 
When there is no evidence collected guiding to any of the $K$ classes, the beliefs reach their minimal values $0$, while the overall uncertainty reaches its maximal value $1$. In practice, DST can be applied to Deep Learning models by fitting a Dirichlet distribution on the model's outputs, in place of the standard categorical distribution. Additionally, the Bayes-Risk loss function \citep{sensoy2018evidential} is used to train the model in replacement of the standard cross-entropy loss. 

\subsection{Other UQ methods}
Finally, we found a few methods not conforming to any of the frameworks previously introduced. We list these applications in \ref{table:others}, with a short description of each UQ approach proposed. 

\section{How to evaluate uncertainty quantification approaches} \label{sec:eval}
\begin{figure*}[ht!]
\centering
\includegraphics[width=0.9\textwidth]{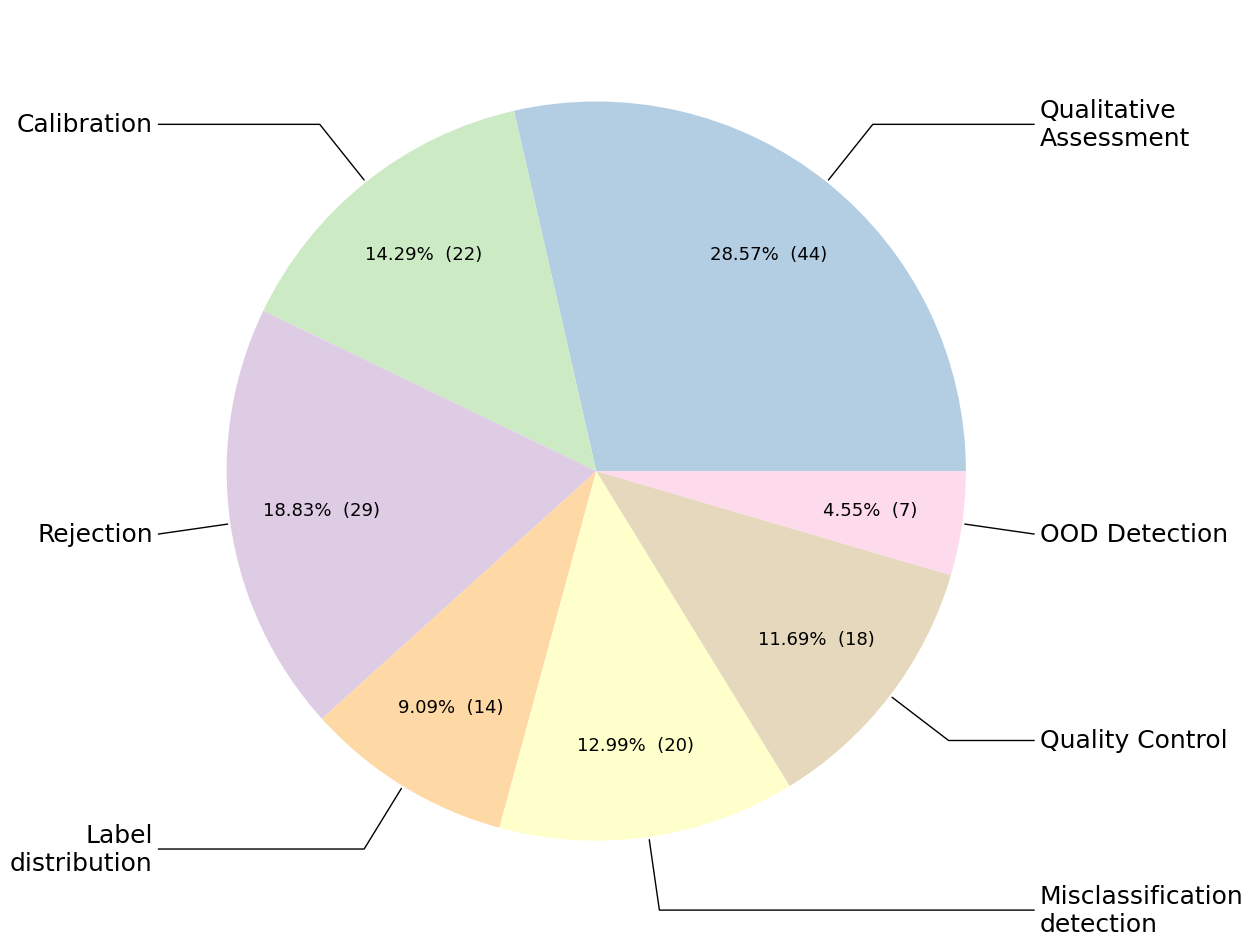}
\caption[]{Implemented UQ evaluation protocols in the reviewed papers. The percentage (and the number) of the reviewed papers per class is mentioned in the Pie chart.}
\label{fig:metrics_pie}
\end{figure*}

In the previous section, we have presented the main UQ approaches that are applied to DL-based medical image classification and segmentation. In this section, we now propose to introduce the different protocols that are implemented in these papers to evaluate the relevance of the UQ approaches. Evaluating UQ approaches is not straight forward, as we typically do not dispose of ground-truth uncertainty values. Proxy metrics are thus developed to estimate the performances of uncertainty quantification methods. We have identified 7 types of evaluation protocols (see Figure \ref{fig:metrics_pie}). In the following, we present each protocol and identify their use cases. Table \ref{table:eval} lists use cases of each metrics in the reviewed corpus of papers.

\subsection{Qualitative assessment protocol}
As computing quantitative metrics for uncertainty is not direct, several works focused on a qualitative assessment of the computed uncertainty estimates. In this context, a visual inspection of the cases considered as certain/uncertain is usually performed to verify whether they correspond to cases that a human would consider as uncertain. Alternatively, the pertinence of the incorporation of UQ in a medical image processing pipeline can be assessed via the monitoring of its beneficial impact on a downstream task (\textit{e.g.} training-image selection in a semi-supervised learning, or improvement of the predictive performance). 

\subsection{Calibration protocol}
As presented in Section \ref{section:softmax_methods}, the output softmax probabilities of a NN can directly be used as a marker of (un)certainty. A popular way of estimating the accuracy of such uncertainty estimates is the use of calibration metrics, that verify the correspondence between predicted probabilities and error rates. Usual choices consist of the Expected Calibration Error (ECE) \citep{guo2017calibration}, the Brier Score, or the Negative Log-Likelihood (NLL) score. 

\subsection{Misclassification detection protocol}\label{section:misclassification}
\begin{figure}[!htb]
\centering
\includegraphics[width=0.5\textwidth]{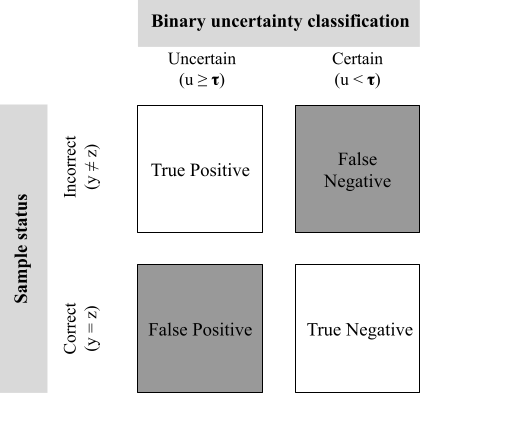}
\caption{Confusion matrix for uncertainty-based misclassification detection. Desired cases are represented in white, while undesired cases are presented in gray. } \label{conf_matrix_miscla}
\end{figure}

A direct downstream application of uncertainty in an automated pipeline is the detection of samples for which the prediction is likely to be incorrect. This is crucial to prevent silent errors that could have dramatic impact, especially in real-world medical image applications. In that sense, the uncertainty estimates can be turned into a binary classifier that aims at distinguishing between correct and incorrect predictions (\textit{i.e.}, sample for which the predicted label $y$ and the ground truth label $z$ differ). As in the binary classification setting, an uncertainty threshold is applied to distinguish between positive (\textit{i.e.} certain) and negative (\textit{i.e.} uncertain) samples. The result of this classification is then compared to the true label of each sample, namely correct or incorrect. In that context, a confusion matrix from the uncertainty point of view can be constructed, by distinguishing 4 possibles cases, as shown in Figure \ref{conf_matrix_miscla}. Usual classification metrics can then be computed based on the counts of True Positive (TP): the classification is uncertain and the expected label and the prediction differ, False Negative (FP): the classification is certain but the expected label and the prediction differ, True Negative (TN): the classification is certain and the expected label and the prediction are identical, and False Negative (FN): the classification is uncertain but the prediction and the expected label are identical. 

\subsection{Rejection protocol}
Another way of exploiting uncertainty estimates in an automated pipeline is the rejection mechanism. In this context, predictions of the model are ordered from the most certain to the most uncertain. A fraction of the most uncertain predictions are then rejected, and the performance of the model is computed on the remaining predictions. If uncertainty estimates efficiently identify uncertain cases that are more likely to be incorrect, then the performance and the remaining prediction should improve. Multiple fractions can be used, producing a curve showing the performance of the model with respect to the fraction of rejected data. The area under the resulting curve is used as a qualitative score. This rejection-based evaluation protocol essentially highlights the same trends as the previous misclassification detection setting. 

\subsection{OOD detection protocol}
A desired property of uncertainty is to be high for abnormal images that are different from the images seen during training. Similarly to the misclassification detection setting, the uncertainty estimates can be translated into a binary classifier that aim at distinguishing between in-distribution (ID) and OOD images. Standard classification metrics can further be computed from the confusion matrix, as presented above in Section \ref{section:misclassification}. 

\subsection{Quality-control protocol}
For segmentation tasks, uncertainty estimates are obtained for every pixel (or voxel) in the medical image. These scores can then be aggregated into image-wise uncertainty, for instance by taking their mean. The correlation between this image-wise uncertainty, and image-wise metrics quantifying the quality of the segmentation, such as the Dice score, can be computed. In an automated medical image segmentation pipeline, this process can be used to detect images for which the produced segmentation does not meet quality standards. We refer to this mode of evaluation, specific to segmentation tasks, as QC-based evaluation protocols. 

\subsection{Label-distribution protocol}
Finally, label-distribution protocol consists in comparing the predicted distribution of labels $P_{out}$ with the ground truth distribution of the experts $P_{gt}$. A popular choice of metric is the Generalized Energy Distance \citep{kohl2018probabilistic} between both distributions.

\section{Discussion} \label{sec:discussion}
We reviewed the most popular UQ methods for DL-based medical image analysis and the associated evaluation protocols. In this section, we list the keys insights of this review, and identify potential future research directions. 

First, the large number of studies incorporating UQ in their medical analysis pipeline proves that the need for UQ is well taken into account by the DL community. This shows that efforts are being made to develop AI tools that are not only powerful, but also useful in a real clinical setting. In this context, the predictive performance of the model only is not enough to reach a good acceptance. UQ is key to facilitate the human-machine collaboration and break the black-box effect.

Bayesian methodology, although providing a strong theoretical background for uncertainty, is scarcely implemented for medical image analysis. This can be explained by the complex implementation that requires (i) the modification of the NN (weights are replaced by distributions, thus doubling the number of parameters to estimate) and (ii) the modification of the training paradigm. Additionally, convergence tends to be slow for complex scenarios \citep{osawa2019practical} and gradient descent more noisy and unstable using a Bayesian NN than with a standard non-deterministic NN \citep{jospin2022hands}. Finally, it has also been observed that Bayesian NNs tend to underfit \citep{dusenberry2020efficient} and that their predictive performances are lower than standard NNs \citep{wenzel2020good}. Approximations of the Bayesian framework, such as dropout-based methods, are thus generally preferred.
    
Overall, MC dropout method seems to be the most popular approach for UQ in medical image analysis, representing nearly half of the implemented methods ($44.73\%$, considering both the standard MC dropout methods ($41.71\%$) as well as sampling Heteroscedastic models ($3.02\%$), which are an MC-dropout extension). This popularity can be explained by its easy implementation in any NN trained with dropout, indeed a large majority of NNs. Additionally, dropout helps preventing over-fitting during training, which is a common problem in medical domain, where training-dataset size is limited.  However, the performance of MC dropout is highly dependent on the applied dropout rate \citep{osawa2019practical}, which can makes it impractical to tune. Moreover, it requires multiple inferences for the same input image, considerably extending the inference time, which may not be compatible with AI applications in clinical environments. 
    
Ensembling approaches are also commonly employed for UQ, although less common than MC dropout models. Aggregating the predictions of multiple models is a popular trick to improve the predictive performance, while also providing quality uncertainty estimates. The drawback is an increased computational cost and time as it requires multiple training and their predictions aggregation at testing.
    
In the literature, a large variety of evaluation protocols are reported, aiming at assessing the quality of uncertainty estimates. In the context of medical image segmentation, if multiple manual expert delineations are available for a given input image, the inter-rater variability can be used as ground truth uncertainty, to be compared with the predicted one. However, most of the time, the corresponding uncertainty values are not provided. Thus, evaluation of UQ usually relies on proxy tasks, such as the detection of misclassification, Out-of-distribution, or Quality Control. These methods are inspired from concrete applications of uncertainty in a real-world scenario. Yet, although commonly used, UQ evaluation based on misclassification detection should not directly be used for ranking methods. Indeed, the set of correct and incorrect predictions is specific to each model that produces its own binary misclassification. It is then inappropriate to compare them directly \citep{ashukha2020pitfalls}. 
    
In this review, we have distinguished between sampling and single-step UQ methods. The latter approaches offer to compute uncertainty in a quick and efficient manner, which is generally required for medical applications. Although they currently represent a niche, their easy practical considerations may promote their rapid adoption. These methods, however, only partially model uncertainty, by computing either aleatoric or epistemic uncertainties, but not both.
    
Finally, it must be acknowledged that the effort of the community is promoted by challenges, such as the 2020 edition of the BraTS challenge \citep{mehta2022qu,menze2014multimodal} that included an UQ task, the MICCAI QUBIQ challenge\footnote{https://qubiq21.grand-challenge.org/} that focused on label uncertainty, and the SHIFT 2022 challenge\footnote{https://shifts.grand-challenge.org/}  that will contain a task of uncertainty quantification for Multiple Sclerosis lesions segmentation \citep{malinin2022shifts}.
    
\subsection{Future directions}
Based on the gaps identified above, we suggest several future research directions for UQ in DL-based medical image analysis. 

As shown, the vast majority ($81.15\%$) of the implemented UQ methods are based on a sampling protocol, aiming at generating multiple predictions for the same query input. Yet, this significantly increases the computational burden of UQ, which may prevent its adoption in an automated pipeline in medical domain. Deterministic UQ methods requiring a single-step to compute uncertainty, are very promising and should be more intensively explored. 
    
Overall, the detection of Out-of-distribution (OOD) predictions using uncertainty concerns few studies, despite of being crucial in a real-world medical scenarios. In an automated medical image pipeline, input samples can exhibit various anomalies that may disturb the functioning of the NN, thus resulting in very poor predictions. Real clinical cases may include artifacts, present a pathology unseen during training, or an unusual contrast setting due to a particular acquisition protocol. In such situations, uncertainty associated to the computed predictions are expected to be high and should represent a warning for the user (\textit{e.g.}, the medical practitioner). In practice, this is usually not the case with standard approaches such as softmax uncertainty, MC dropout or Deep Ensemble, which have limited performance in terms of OOD detection \citep{ovadia2019can,ulmer2021know}. This motivates the development of feature-based methods, specially tailored for OOD detection. Note that OOD detection is a very active research topic not specific to the UQ field \citep{bulusu2020anomalous}, but currently rarely exploited for medical image analysis. We hypothesize that the lack of OOD-based evaluation protocol may be due to the difficulty of gathering relevant data. A simple solution may be to use two distinct datasets, one for training the neural network and its evaluation on in-distribution data, and one during testing for OOD detection \citep{karimi2020improving}. Another approach would be to use data augmentation to corrupt images with synthetic artifacts, helping to achieve a more realistic setting \citep{shaw2021decoupled}.
     
Finally, while the need for UQ in medical applications is unquestionable, we argue that being able to understand the prediction process of the DL model is also crucial to promote a trustable usage of AI in medicine. Then, the link between explainability and uncertainty should be studied, which would allow to understand both \textit{how} the prediction is made, and whether or not it should be \textit{trusted}. An interesting research direction would be to complement uncertainty estimates with explanations, helping the user to understand the sources of uncertainty in an intelligible way and possibly contribute to its improvement. 
     
\section{Conclusion}
In this review, we have proposed an overview of the most popular UQ methods implemented in DL-based medical image applications, a specific domain with inherent uncertainty. Numerous phenomenons can cause predictive uncertainty, such as noisy images, imperfect ground truth labels, lack or incomplete data, and inter-site image variability. The literature proposes various methods to quantify this uncertainty which are applied to a very large range of medical image applications. As demonstrated in this review, developing trustable AI solutions integrating uncertainty quantification of the computed predictions is an active research topics. 

\section{Declaration of competing interest}
BL, AT, SD, HD are employees of the Pixyl company. MD and FF serve on Pixyl scientific advisory board. 

\section{Acknowledgments}

\label{sec:acknowledgments}
Benjamin Lambert is supported by a CIFRE convention (ANRT 2020/1555). The authors declare that they have no known competing financial interests or personal relationships that could have appeared to influence the work reported in this paper.

\appendix
\section{Classification of the reviewed papers with respect to the UQ methods implemented.} \label{table:methods}

\begin{center}
\begin{small}
\footnotesize
\begin{longtable}{lccc}
\toprule

\textbf{Study} & \textbf{Year} & \textbf{Modality} & \textbf{Applications} \\ \midrule

%%%%%%%%%%%%%%%%%%%%%%%%%%%%%%%%%%%%%%%%%%%%%%%%%%%%%%%%%%%%%%%%%%%%%%%%%%%%%%%
\rowcolor{lightgray}\multicolumn{4}{c}{Softmax} \\ \midrule
%%%%%%%%%%%%%%%%%%%%%%%%%%%%%%%%%%%%%%%%%%%%%%%%%%%%%%%%%%%%%%%%%%%%%%%%%%%%%%%

\cite{wang2018interactive} & 2018 & \begin{tabular}[c]{@{}c@{}}fetal MRI\\ brain MRI\end{tabular} & \begin{tabular}[c]{@{}c@{}}organ segmentation\\ tumor segmentation\end{tabular} \\ \hline
\cite{jungo2020analyzing} & 2020 & brain MRI & tumor segmentation \\ \hline
\cite{mojabi2020tissue} & 2020 & breast US & tissue segmentation \\ \hline
\cite{devries2018leveraging} & 2018 & dermatoscopy & segmentation \\ \hline
\cite{filos2019systematic} & 2019 & retinal images  & classification \\ \hline
\cite{rousseau2021post} & 2020 & \begin{tabular}[c]{@{}c@{}}brain MRI\\ brain CT\end{tabular}  & \begin{tabular}[c]{@{}c@{}}tumor segmentation\\ stroke segmentation\end{tabular} \\ \hline
\cite{mehta2020uncertainty} & 2020 & brain MRI & tumor segmentation\\ \hline
\cite{diao2022unified}    & 2022 & brain MRI & tumor segmentation \\ \hline
\cite{hoebel2020exploration}   & 2020 & lung CT & nodule segmentation \\ \hline
\cite{silva2021using}   & 2021 & MRI / CT & \begin{tabular}[c]{@{}c@{}} segmentation\\ tasks \end{tabular} \\ \hline
\cite{carneiro2020deep} & 2021 & colonoscopy & polyps classification \\ \hline
\begin{tabular}[l]{@{}l@{}}\cite{calderon2021improving-covid}\end{tabular}  & 2021 & chest X-Ray & COVID detection  \\ \hline 
\begin{tabular}[l]{@{}l@{}}\cite{calderon2021improving-mammo}\end{tabular} & 2021 & X-Ray & mammogram classification \\ \hline 
\cite{berger2021confidence} & 2021 & chest X-Ray & disease classification\\ \hline 
\cite{liang2020neural} & 2020 & CT / histology &  classification tasks \\ \hline
\cite{ayhan2020expert} & 2020 & retinal images & disease classification \\ \hline
\cite{belharbi2021deep} & 2020 & histology & cancer cell segmentation \\ \hline
\cite{lin2021variance} & 2021 & CT & multi-organ segmentation \\ \hline
\cite{judge2022crisp} & 2022 & \begin{tabular}[c]{@{}c@{}}US \\lung X-Ray \end{tabular} & \begin{tabular}[c]{@{}c@{}}cardiac segmentation\\lung segmentation \end{tabular} \\\bottomrule

%%%%%%%%%%%%%%%%%%%%%%%%%%%%%%%%%%%%%%%%%%%%%%%%%%%%%%%%%%%%%%%%%%%%%%%%%%%%%%%
\rowcolor{lightgray}\multicolumn{4}{c}{Bayesian Neural Network} \\ \midrule
%%%%%%%%%%%%%%%%%%%%%%%%%%%%%%%%%%%%%%%%%%%%%%%%%%%%%%%%%%%%%%%%%%%%%%%%%%%%%%%

\cite{dhakal2021uncertainty} & 2021 & knee MRI & classification \\ \hline
\cite{filos2019systematic} & 2019 & retinal images & classification \\ \hline
\cite{li2021uncertainty} & 2021 & \begin{tabular}[c]{@{}c@{}}lung CT\\ nasal endoscopy\end{tabular} & lesion segmentation \\ \bottomrule

%%%%%%%%%%%%%%%%%%%%%%%%%%%%%%%%%%%%%%%%%%%%%%%%%%%%%%%%%%%%%%%%%%%%%%%%%%%%%%%
\rowcolor{lightgray}\multicolumn{4}{c}{Monte Carlo dropout} \\ \midrule
%%%%%%%%%%%%%%%%%%%%%%%%%%%%%%%%%%%%%%%%%%%%%%%%%%%%%%%%%%%%%%%%%%%%%%%%%%%%%%%

\cite{jungo2018effect} & 2018 & brain MRI & cavity segmentation \\ \hline
\cite{zhang2022epistemic} & 2022 & \begin{tabular}[c]{@{}c@{}}lung CT\\ abdominal CT\end{tabular} & \begin{tabular}[c]{@{}c@{}}nodule segmentation\\ tumor segmentation\end{tabular} \\ \hline
\cite{ghosal2021uncertainty} & 2021 & biopsy & tumor segmentation \\ \hline
\cite{ghoshal2020estimating} & 2020 & lung CT & COVID detection \\ \hline
\cite{yu2019uncertainty} & 2019 & cardiac MRI & cardiac segmentation \\ \hline
\cite{wickstrom2020uncertainty} & 2018 & colonoscopy & polyps segmentation \\ \hline
\cite{ghoshal2021estimating} & 2021 & \begin{tabular}[c]{@{}c@{}}microscopy\\ brain MRI\end{tabular} & \begin{tabular}[c]{@{}c@{}}nuclei segmentation\\ tumor classification\end{tabular} \\ \hline
\cite{ozdemir20193d} & 2019 & lung CT & nodule segmentation \\ \hline
\cite{sander2019towards} & 2018 & cardiac MRI & segmentation \\ \hline
\cite{eaton2018towards} & 2018 & brain MRI & tumor segmentation \\ \hline
\cite{abdar2021uncertainty} & 2021 & dermatoscopy & classification \\ \hline
\cite{tousignant2019prediction} & 2019 & brain MRI & classification \\ \hline
\cite{jungo2020analyzing} & 2020 & brain MRI & tumor segmentation \\ \hline
\cite{balagopal2021deep} & 2021 & prostate CT & target segmentation \\ \hline
\cite{hu2020coarse} & 2020 & brain CT/PET & tumor segmentation \\ \hline
\cite{xia2020uncertainty} & 2020 & CT & segmentation \\ \hline
\cite{sedai2019uncertainty} & 2019 & retinal image & layer segmentation \\ \hline
\cite{wang2019aleatoric} & 2019 & brain MRI & \begin{tabular}[c]{@{}c@{}}brain segmentation\\ tumor segmentation\end{tabular}       \\ \hline
\cite{mehrtash2020confidence} & 2020 & MRI & segmentation \\ \hline
\cite{roy2019bayesian} & 2018 & brain MRI & brain segmentation \\ \hline
\cite{devries2018leveraging}& 2018 & dermatoscopy & segmentation \\ \hline
\cite{liu2020exploring} & 2020 & prostate MRI & segmentation \\ \hline
\cite{dhakal2021uncertainty} & 2021& knee MRI & classification \\ \hline
\cite{jungo2017towards} & 2017 & brain MRI & tumor segmentation \\ \hline
\cite{rkaczkowski2019ara} & 2019 & microscopy & classification \\ \hline
\cite{filos2019systematic} & 2019 & retinal image & classification \\ \hline
\cite{abideen2020uncertainty} & 2020 & chest CT & tuberculosis detection \\ \hline
\cite{thagaard2020can} & 2020 & histology & metastasis detection \\ \hline
\cite{jungo2018uncertainty} & 2018 & brain MRI & cavity segmentation \\ \hline
\cite{hiasa2019automated} & 2019 & muscle CT & muscle segmentation \\ \hline
\cite{orlando2019u2} & 2019 & retinal images & layer segmentation \\ \hline
\cite{leibig2017leveraging} & 2017 & retinal images & classification \\ \hline
\cite{kwon2020uncertainty} & 2020 & \begin{tabular}[c]{@{}c@{}}retinal images\\ brain MRI\end{tabular} & \begin{tabular}[c]{@{}c@{}}vessel segmentation\\ stroke segmentation\end{tabular} \\ \hline
\cite{rousseau2021post} & 2020 & \begin{tabular}[c]{@{}c@{}}brain MRI \\ brain CT\end{tabular} & \begin{tabular}[c]{@{}c@{}}tumor segmentation\\ stroke segmentation\end{tabular} \\ \hline
\cite{mehta2020uncertainty} & 2020 & brain MRI & tumor segmentation \\ \hline
\begin{tabular}[l]{@{}l@{}}\cite{asgharnezhad2022objective}\end{tabular} & 2022 & lung CT & COVID detection \\ \hline
\cite{yang2021exploring} & 2020 & lung CT & nodule detection \\ \hline
\cite{ozdemir2017propagating} & 2017 & lung CT & nodule detection \\ \hline
\begin{tabular}[l]{@{}l@{}}\cite{soberanis2020uncertainty}\end{tabular} & 2019 & CT & segmentation \\ \hline
\cite{pan2019prostate} & 2019 & prostate MRI & segmentation \\ \hline
\cite{lee2022method} & 2022 & brain MRI & tumor segmentation \\ \hline
\cite{bhat2021using} & 2021 & liver MRI & metastase segmentation \\ \hline
\cite{huang2020heterogeneity} & 2020 & cardiac OCT & tissue segmentation \\ \hline
\cite{iwamoto2021improving} & 2021 & microscopy & segmentation \\ \hline
\cite{mcclure2019knowing} & 2019 & brain MRI & atlas segmentation \\ \hline
\cite{natekar2020demystifying} & 2019 & brain MRI & tumor segmentation \\ \hline
\cite{herzog2020integrating} & 2020 & brain MRI & stroke classification \\ \hline
\cite{mehta2019propagating} & 2019 & brain MRI & \begin{tabular}[c]{@{}c@{}}MS lesion segmentation\\ tumor segmentation\end{tabular} \\ \hline
\cite{molle2019quantifying} & 2019 & dermatoscopy & classification \\ \hline
\cite{zou2022tbrats} & 2022 & brain MRI & tumor segmentation \\ \hline
\cite{diao2022unified} & 2022 & brain MRI & tumor segmentation \\ \hline
\cite{hoebel2020exploration} & 2020 & lung CT & nodule segmentation \\ \hline
\cite{karimi2020improving} & 2020 & MRI / CT  & segmentation \\ \hline
\cite{redekop2021uncertainty} & 2020 & \begin{tabular}[c]{@{}c@{}}CT\\ dermatoscopy\end{tabular} & segmentation \\ \hline
\cite{bhat2022extending} & 2021 & MRI/CT  &  segmentation tasks \\ \hline
\cite{song2021bayesian} & 2021 & tongue images & oral cancer classification \\ \hline
\cite{gou2021deep} & 2021 & head CT & \begin{tabular}[c]{@{}c@{}}detection of \\ subarachnoid hemorrhages  \end{tabular}\\ \hline
\cite{carneiro2020deep} & 2021 & colonoscopy & polyps classification \\ \hline
\cite{mahapatra2021interpretability} & 2021 & \begin{tabular}[c]{@{}c@{}}chest X-Ray\\ histology\end{tabular} & \begin{tabular}[c]{@{}c@{}}disease classification\\ gland segmentation \end{tabular} \\ \hline 
\cite{hasan2021multi} & 2022 & cardiac MRI & segmentation \\ \hline
\cite{mojiri2022deep} & 2022 & brain MRI & \begin{tabular}[c]{@{}c@{}}White Matter\\ Hyperintensities segmentation \end{tabular} \\ \hline
\cite{laves2019uncertainty} & 2019 & retina images & disease classification \\ \hline
\cite{mobiny2019risk} & 2019 & skin images & disease classification \\ \hline
 \cite{calderon2021improving-covid}  & 2021 & chest X-Ray & COVID detection  \\ \hline 
\cite{calderon2021improving-mammo} & 2021 & X-Ray & mammogram classification \\ \hline 
\cite{linmans2020efficient} & 2020 & microscopy & \begin{tabular}[c]{@{}c@{}}segmentation of breast\\ cancer metastasis \end{tabular} \\ \hline
\cite{berger2021confidence} & 2021 & chest X-Ray & disease classification\\ \hline 
\cite{ayhan2020expert} & 2020 & retina images & disease classification \\ \hline
\cite{ju2022improving} & 2021 & dermatoscopy & disease classification \\ \hline
\cite{hasan2022calibration} & 2022 & cardiac MRI & segmentation \\ \hline
\cite{jimenez2022curriculum} & 2020 & femur X-Ray & fracture classification \\ \hline
\cite{cao2021dilated} & 2019 & breast ultrasound & breast mass segmentation \\ \hline
\cite{pocevivciute2022generalisation} & 2022 & microscopy & cancer classification \\ \hline
\cite{rajaraman2022uncertainty} & 2022 & chest X-Ray & \begin{tabular}[c]{@{}c@{}}tuberculosis consistent\\ region segmentation \end{tabular} \\ \hline
\cite{senousy2021mcua} & 2021 & histology & breast cancer classification \\ \hline
\cite{ahsan2022active} & 2022 & retina images & disease classification \\ \hline
\cite{javadi2022towards} & 2022 & prostate ultrasound & cancer detection \\ \hline 
\cite{tardy2019uncertainty} & 2019 & X-Ray & \begin{tabular}[c]{@{}c@{}}mammogram\\ classification \end{tabular} \\ \hline
\cite{yang2021uncertainty} & 2021 & CT / histology & disease classification \\ \hline 
\cite{jensen2019improving} & 2019 & dermatoscopy & disease classification \\ \hline
\cite{lambert2022beyond} & 2022 & brain MRI & MS lesions segmentation \\ \hline
\cite{judge2022crisp} & 2022 & \begin{tabular}[c]{@{}c@{}}US \\lung X-Ray \end{tabular} & \begin{tabular}[c]{@{}c@{}}cardiac segmentation\\lung segmentation \end{tabular} \\\hline
\cite{zhao2022efficient} & 2022 & MRI & cardiac segmentation \\ \bottomrule

%%%%%%%%%%%%%%%%%%%%%%%%%%%%%%%%%%%%%%%%%%%%%%%%%%%%%%%%%%%%%%%%%%%%%%%%%%%%%%%
\rowcolor{lightgray}\multicolumn{4}{c}{Ensemble} \\ \midrule
%%%%%%%%%%%%%%%%%%%%%%%%%%%%%%%%%%%%%%%%%%%%%%%%%%%%%%%%%%%%%%%%%%%%%%%%%%%%%%%

\cite{yang2017suggestive} & 2017 & \begin{tabular}[c]{@{}c@{}}histology \\ US\end{tabular} & \begin{tabular}[c]{@{}c@{}} gland / lymph nodes \\ segmentation\end{tabular} \\ \hline
\cite{abdar2021uncertainty}  & 2021 & dermatoscopy & cancer classification \\ \hline
\cite{jungo2020analyzing}  & 2020 & brain MRI & tumor segmentation \\ \hline
\cite{shamsi2021uncertainty} & 2021 & lung CT & COVID detection \\ \hline
\cite{mehrtash2020confidence} & 2020 & MRI &  segmentation tasks \\ \hline
\cite{filos2019systematic}  & 2019 & retinal images & classification \\ \hline
\cite{thagaard2020can} & 2020 & histology & metastasis detection \\ \hline
\cite{mehta2020uncertainty} & 2020 & brain MRI & tumor segmentation \\ \hline
\begin{tabular}[l]{@{}l@{}}\cite{asgharnezhad2022objective}\end{tabular} & 2022 & lung CT & COVID detection \\ \hline
\cite{vu2020multi} & 2020 & brain MRI & tumor segmentation \\ \hline
\cite{zhou2021survey}  & 2022 & brain MRI & tumor segmentation \\ \hline
\cite{yang2022uncertainty} & 2022 & CT / MRI &  segmentation tasks \\ \hline
\cite{hoebel2020exploration} & 2020 & lung CT & nodule segmentation \\ \hline
\cite{mehrtash2021prostate} & 2021 & prostate MRI & cancer classification \\ \hline
\cite{redekop2021uncertainty} & 2021 & \begin{tabular}[c]{@{}c@{}}liver CT\\ dermatoscopy \end{tabular} & segmentation
\\ \hline
\cite{cetindag2022meta} & 2021 & MRI/CT &  segmentation tasks \\ \hline
\cite{pal2022holistic} & 2021 & MRI/CT &  segmentation tasks \\ \hline
\cite{wang2020ud} & 2020 & retinal images & disease classification \\ \hline
\cite{linmans2020efficient} & 2020 & microscopy & \begin{tabular}[c]{@{}c@{}}segmentation of breast\\cancer metastasis\end{tabular} \\ \hline
\cite{berger2021confidence} & 2021 & chest X-Ray & disease classification\\ \hline
\cite{ghesu2019quantifying} & 2019 & Chest X-Ray & classification \\ \hline
\cite{ghesu2021quantifying} & 2019 &
\begin{tabular}[c]{@{}c@{}}Chest X-Ray \\ brain MRI \\ abdominal US \end{tabular} & \begin{tabular}[c]{@{}c@{}}disease classification\\ detection of metastases \\ view classification  \end{tabular} \\ \hline
\cite{ayhan2020expert} & 2020 & retinal images & disease classification \\ \hline
\cite{guo2022cardiac} & 2022 & cardiac MRI & segmentation \\ \hline 
\begin{tabular}[l]{@{}l@{}}\cite{rosas2021asymmetric}\end{tabular} & 2021 & brain MRI & tumor segmentation \\ \hline 
\cite{pocevivciute2022generalisation} & 2022 & microscopy & cancer classification \\ \hline
\cite{yang2021uncertainty} & 2021 & CT / histology & disease classification \\ \hline 
\cite{jensen2019improving} & 2019 & dermatoscopy & disease classification \\ \hline
\cite{xiang2022fussnet} & 2022 & \begin{tabular}[c]{@{}c@{}}CT \\ MRI \end{tabular} & \begin{tabular}[c]{@{}c@{}}pancreas segmentation \\ cardiac segmentation \end{tabular} \\ \hline
\cite{kushibar2022layer} & 2022 & mammogram & mass segmentation \\ \hline 
\cite{zhao2022efficient} & 2022 & MRI & cardiac segmentation \\ \bottomrule

%%%%%%%%%%%%%%%%%%%%%%%%%%%%%%%%%%%%%%%%%%%%%%%%%%%%%%%%%%%%%%%%%%%%%%%%%%%%%%%
\rowcolor{lightgray}\multicolumn{4}{c}{Monte Carlo dropout Ensemble} \\ \midrule
%%%%%%%%%%%%%%%%%%%%%%%%%%%%%%%%%%%%%%%%%%%%%%%%%%%%%%%%%%%%%%%%%%%%%%%%%%%%%%%

\cite{ghoshal2021estimating} & 2021 & \begin{tabular}[c]{@{}c@{}}microscopy\\ brain MRI\end{tabular} & \begin{tabular}[c]{@{}c@{}}nuclei segmentation\\ tumor classification\end{tabular} \\ \hline
\cite{abdar2021uncertainty}  & 2021 & dermatoscopy  & cancer classification \\ \hline
\cite{filos2019systematic}  & 2019 & retinal images & classification \\ \hline
\cite{mehta2020uncertainty} & 2020 & brain MRI & tumor segmentation \\ \hline
\cite{asgharnezhad2022objective} & 2022 & lung CT & COVID detection \\ \hline
\cite{abdar2021uncertaintyfusenet} & 2022 & chest X-RAY / CT & COVID detection \\ \hline \cite{yang2021uncertainty} & 2021 & CT / histology & disease classification \\ \bottomrule

%%%%%%%%%%%%%%%%%%%%%%%%%%%%%%%%%%%%%%%%%%%%%%%%%%%%%%%%%%%%%%%%%%%%%%%%%%%%%%%
\rowcolor{lightgray}\multicolumn{4}{c}{Heteroscedastic (Sampling)} \\ \midrule
%%%%%%%%%%%%%%%%%%%%%%%%%%%%%%%%%%%%%%%%%%%%%%%%%%%%%%%%%%%%%%%%%%%%%%%%%%%%%%%

\cite{nair2020exploring} & 2020 & brain MRI  & MS lesion segmentation \\ \hline
\cite{eaton2018towards} & 2018 & brain MRI & tumor segmentation \\ \hline
\cite{jungo2020analyzing} & 2020 & brain MRI & tumor segmentation \\ \hline
\cite{sedai2018joint} & 2018 & retinal images & retinal layer segmentation \\ \hline
\cite{devries2018leveraging} & 2018 & dermatoscopy & segmentation \\ \hline
\cite{shaw2021decoupled} & 2021 & brain MRI & brain segmentation \\ \bottomrule
 
%%%%%%%%%%%%%%%%%%%%%%%%%%%%%%%%%%%%%%%%%%%%%%%%%%%%%%%%%%%%%%%%%%%%%%%%%%%%%%%
\rowcolor{lightgray}\multicolumn{4}{c}{Heteroscedastic (Deterministic)} \\ \midrule
%%%%%%%%%%%%%%%%%%%%%%%%%%%%%%%%%%%%%%%%%%%%%%%%%%%%%%%%%%%%%%%%%%%%%%%%%%%%%%%

\cite{mckinley2020uncertainty} & 2020 & brain MRI & tumor segmentation \\ \hline
\cite{mckinley2020automatic} & 2020 & brain MRI & \begin{tabular}[c]{@{}c@{}}new MS\\lesion segmentation\end{tabular} \\ \hline
\cite{mckinley2018ensembles} & 2018 & brain MRI & tumor segmentation \\ \hline
\cite{mckinley2019triplanar} & 2019 & brain MRI & tumor segmentation \\ \hline
\cite{devries2018leveraging} & 2018 & dermatoscopy & segmentation \\ \hline
\cite{diao2022unified} & 2022 & brain MRI & tumor segmentation \\ \hline
\cite{judge2022crisp} & 2022 & \begin{tabular}[c]{@{}c@{}}US \\lung X-Ray \end{tabular} & \begin{tabular}[c]{@{}c@{}}cardiac segmentation\\lung segmentation \end{tabular} \\ \bottomrule

%%%%%%%%%%%%%%%%%%%%%%%%%%%%%%%%%%%%%%%%%%%%%%%%%%%%%%%%%%%%%%%%%%%%%%%%%%%%%%%
\rowcolor{lightgray}\multicolumn{4}{c}{Label Distribution models} \\ \midrule
%%%%%%%%%%%%%%%%%%%%%%%%%%%%%%%%%%%%%%%%%%%%%%%%%%%%%%%%%%%%%%%%%%%%%%%%%%%%%%%

\cite{kohl2018probabilistic} & 2018 & lung CT & nodule segmentation \\ \hline
\cite{kohl2019hierarchical} & 2019 & \begin{tabular}[c]{@{}c@{}}lung CT\\ microscopy\end{tabular} & \begin{tabular}[c]{@{}c@{}}nodule segmentation\\ neocortex segmentation\end{tabular} \\ \hline
\cite{baumgartner2019phiseg} & 2019 & \begin{tabular}[c]{@{}c@{}}lung CT\\ prostate MRI\end{tabular}  & \begin{tabular}[c]{@{}c@{}}nodule segmentation\\ prostate segmentation\end{tabular} \\ \hline
\cite{hu2019supervised} & 2019 & \begin{tabular}[c]{@{}c@{}}lung CT\\ prostate MRI\end{tabular} & \begin{tabular}[c]{@{}c@{}}nodule segmentation\\ prostate segmentation\end{tabular} \\ \hline
\cite{li2020uncertainty} & 2020 & prostate MRI & prostate segmentation \\ \hline
\cite{gantenbein2020revphiseg} & 2020 & \begin{tabular}[c]{@{}c@{}}lung CT\\ prostate MRI\end{tabular} & \begin{tabular}[c]{@{}c@{}}nodule segmentation\\ prostate segmentation\end{tabular} \\ \hline
\cite{monteiro2020stochastic} & 2020 & \begin{tabular}[c]{@{}c@{}}lung CT\\ brain MRI\end{tabular} & \begin{tabular}[c]{@{}c@{}}nodule segmentation\\ tumor segmentation\end{tabular} \\ \hline
\cite{zou2022tbrats} & 2022 & brain MRI & tumor segmentation \\ \hline
\cite{diao2022unified} & 2022 & brain MRI & tumor segmentation \\ \hline
\cite{selvan2020uncertainty} & 2020 & \begin{tabular}[c]{@{}c@{}}lung CT\\ retinal image\end{tabular} & \begin{tabular}[c]{@{}c@{}}nodule segmentation\\ vessel segmentation\end{tabular} \\ \hline
\cite{cetindag2022meta} & 2021 & MRI/CT  &  segmentation tasks \\ \hline
\cite{ji2020uncertainty} & 2020 & MRI/CT &  segmentation tasks \\ \hline
\cite{bhat2022extending} & 2021 & MRI/CT &  segmentation tasks \\ \bottomrule

%%%%%%%%%%%%%%%%%%%%%%%%%%%%%%%%%%%%%%%%%%%%%%%%%%%%%%%%%%%%%%%%%%%%%%%%%%%%%%%
\rowcolor{lightgray}\multicolumn{4}{c}{Test-Time Augmentation} \\ \midrule
%%%%%%%%%%%%%%%%%%%%%%%%%%%%%%%%%%%%%%%%%%%%%%%%%%%%%%%%%%%%%%%%%%%%%%%%%%%%%%%

\cite{wang2019aleatoric} & 2019 & brain MRI & \begin{tabular}[c]{@{}c@{}}fetal brain segmentation\\ tumor segmentation\end{tabular} \\ \hline
\cite{norouzi2019exploiting} & 2019 & cardiac MRI & cardiac segmentation \\ \hline
\cite{ayhan2018test} & 2018 & retinal images & classification \\ \hline
\cite{pan2019prostate} & 2019 & prostate MRI & prostate segmentation \\ \hline
\cite{diao2022unified} & 2022 & brain MRI & tumor segmentation \\ \hline
\cite{redekop2021uncertainty} & 2021 & \begin{tabular}[c]{@{}c@{}}liver CT\\ dermatoscopy \end{tabular} & segmentation \\ \hline 
\cite{combalia2020uncertainty} & 2020 & dermatoscopy & lesion classification \\ \hline
\cite{ayhan2020expert} & 2020 & retinal images & disease classification \\ \hline
\cite{wang2019automatic} & 2019 & brain MRI & tumor segmentation \\ \hline 
\cite{pocevivciute2022generalisation} & 2022 & microscopy & cancer classification \\ \hline
\cite{javadi2022towards} & 2022 & prostate US & cancer detection \\ \hline 
\cite{jensen2019improving} & 2019 & dermatoscopy & disease classification \\ \bottomrule

%%%%%%%%%%%%%%%%%%%%%%%%%%%%%%%%%%%%%%%%%%%%%%%%%%%%%%%%%%%%%%%%%%%%%%%%%%%%%%%
\rowcolor{lightgray}\multicolumn{4}{c}{Feature-based methods} \\ \midrule
%%%%%%%%%%%%%%%%%%%%%%%%%%%%%%%%%%%%%%%%%%%%%%%%%%%%%%%%%%%%%%%%%%%%%%%%%%%%%%%

\cite{karimi2020improving} & 2020 & CT, MRI & \begin{tabular}[c]{@{}c@{}}\\ segmentation tasks \end{tabular} \\ \hline
\cite{diao2022unified} & 2022 & brain MRI & tumor segmentation \\ \hline
\begin{tabular}[l]{@{}l@{}}\cite{calderon2021improving-covid}\end{tabular} & 2021 & chest X-Ray & COVID detection  \\ \hline 
\cite{berger2021confidence} & 2021 & chest X-Ray & disease classification\\ \hline 
\cite{tardy2019uncertainty} & 2019 & X-Ray & \begin{tabular}[c]{@{}c@{}}mammogram\\ classification \end{tabular} \\ \bottomrule

%%%%%%%%%%%%%%%%%%%%%%%%%%%%%%%%%%%%%%%%%%%%%%%%%%%%%%%%%%%%%%%%%%%%%%%%%%%%%%%
\rowcolor{lightgray}\multicolumn{4}{c}{Evidential Deep Learning} \\ \midrule
%%%%%%%%%%%%%%%%%%%%%%%%%%%%%%%%%%%%%%%%%%%%%%%%%%%%%%%%%%%%%%%%%%%%%%%%%%%%%%%

\cite{ghesu2019quantifying} & 2019 & chest X-Ray & classification \\ \hline
\cite{ghesu2021quantifying} & 2019 &
\begin{tabular}[c]{@{}c@{}}chest X-Ray \\ brain MRI \\ Abdominal US \end{tabular} & \begin{tabular}[c]{@{}c@{}}disease classification\\ detection of metastases \\ view classification  \end{tabular} \\ \hline
\cite{tardy2019uncertainty} & 2019 & X-Ray & mammogram classification \\ \hline
\cite{huang2021evidential} & 2021 & PET / CT & lymphomas segmentation \\ \hline
\cite{zou2022tbrats} & 2020 & brain MRI & tumor segmentation \begin{tabular}[c]{@{}c@{}}\end{tabular}\\ \bottomrule

\end{longtable}
\end{small}
\end{center}

\section{Uncategorized approaches for UQ in medical image processing applications.} \label{table:others}

\begin{table}[H]
\resizebox{\textwidth}{!}{
\begin{tabular}{@{}lccc@{}}
\toprule
\textbf{Study} & \textbf{Year} & \textbf{Modality} & \textbf{Application} \\ \midrule
\cite{jungo2020analyzing}                     & 2020 & brain MRI  & \begin{tabular}[c]{@{}c@{}}Training of an auxiliary net to predict the voxel-wise \\ errors of a brain tumor segmentation model. \end{tabular} \\ \hline
\cite{mishra2021objective}                   & 2021 & retina images & \begin{tabular}[c]{@{}c@{}} Modeling task-dependent (homoscedastic) uncertainty \\ in a multi-tasking vessel segmentation setting. \end{tabular}         \\ \hline
\cite{follmer2022active} & 2021 & heart CT & \begin{tabular}[c]{@{}c@{}} Modeling task-dependent (homoscedastic) uncertainty \\ in a multi-tasking heart-segmentation setting. \end{tabular} \\ \hline
\cite{laves2019uncertainty} & 2019 & retina images & \begin{tabular}[c]{@{}c@{}} Approximating the output posterior distribution of the \\ classification network by a normal distribution $\mathcal{N}(\mu, \sigma^2)$ \\ and learning both parameters using a variational network. \end{tabular} \\ \hline
\cite{toledo2020hybrid} & 2020 & retina images & \begin{tabular}[c]{@{}c@{}} Addition of a Gaussian Process at the end of a DL \\ model to quantify classification uncertainty. \end{tabular} \\ \hline
\cite{jensen2019improving} & 2019 & dermatoscopy & \begin{tabular}[c]{@{}c@{}}Use of Monte Carlo Batch Normalization (MCBN) \\ to quantify uncertainty by relying on the \\ stochasticity of Batch Normalization layers. \end{tabular} \\ \hline
\cite{judge2022crisp} & 2022 & \begin{tabular}[c]{@{}c@{}}US \\lung X-Ray\end{tabular} & \begin{tabular}[c]{@{}c@{}} Use of Contrastive Learning to model the joint \\distribution of valid segmentations and associated images. \end{tabular} \\ \hline
\cite{lu2022improving} & 2022 & MRI & \begin{tabular}[c]{@{}c@{}}Use Conformal Predictions to provide a set of \\plausible classes for a given image with coverage guarantees.\end{tabular} \\ \bottomrule
\end{tabular}}
\end{table}

\section{Evaluation protocols proposed in the corpus of papers and classified according to their framework.} \label{table:eval}

\newcolumntype{P}[1]{>{\centering\arraybackslash}p{#1}}
\setlength\LTcapwidth{\textwidth}
\afterpage{
\begin{small}
\begin{center}
\begin{longtable}[H]{lP{9cm}}
\toprule
\textbf{Evaluation protocol} & \textbf{Papers} \\ \midrule 
\begin{tabular}[l]{@{}l@{}}
\textbf{Qualitative} \\ \textbf{Assessment} \end{tabular} & \cite{wang2018interactive}, \cite{yu2019uncertainty}, \cite{wickstrom2020uncertainty}, \cite{eaton2018towards}, \cite{mojabi2020tissue}, \cite{shamsi2021uncertainty}, \cite{sedai2018joint}, \cite{norouzi2019exploiting}, \cite{hu2020coarse}, \cite{xia2020uncertainty}, \cite{sedai2019uncertainty}, \cite{li2020uncertainty}, \cite{dhakal2021uncertainty}, \cite{jungo2017towards}, \cite{kwon2020uncertainty}, \cite{mishra2021objective}, \cite{soberanis2020uncertainty}, \cite{lee2022method}, \cite{bhat2021using}, \cite{huang2020heterogeneity}, \cite{mckinley2018ensembles}, \cite{mckinley2019triplanar}, \cite{mckinley2020automatic}, \cite{natekar2020demystifying}, \cite{mehta2019propagating}, \cite{li2021uncertainty}, \cite{mehrtash2021prostate}, \cite{redekop2021uncertainty}, \cite{wang2020ud}, \cite{mahapatra2021interpretability}, \cite{hasan2021multi}, \cite{mojiri2022deep}, \cite{laves2019uncertainty}, \cite{toledo2020hybrid}, \cite{lin2021variance},\cite{guo2022cardiac}, \cite{ju2022improving},  \cite{belharbi2021deep},\cite{hasan2022calibration}, \cite{jimenez2022curriculum}, \cite{cao2021dilated},\cite{senousy2021mcua}, \cite{huang2021evidential},\cite{xiang2022fussnet} \\ \midrule
\textbf{Calibration}          & \cite{ghoshal2021estimating}, \cite{ozdemir20193d}, \cite{sander2019towards}, \cite{jungo2020analyzing}, \cite{mehrtash2020confidence}, \cite{thagaard2020can}, \cite{rousseau2021post},  \cite{asgharnezhad2022objective}, \cite{ozdemir2017propagating},\cite{karimi2020improving}, \cite{zou2022tbrats}, \cite{herzog2020integrating},\cite{gou2021deep}, \cite{carneiro2020deep}, \cite{berger2021confidence},\cite{liang2020neural},\cite{ayhan2020expert}, \cite{javadi2022towards}, \cite{jensen2019improving},\cite{judge2022crisp},\cite{kushibar2022layer},\cite{zhao2022efficient} \\ \midrule
\begin{tabular}[l]{@{}l@{}}
\textbf{Misclassification} \\ \textbf{detection} \end{tabular} &  \cite{ghoshal2020estimating}, \cite{yang2017suggestive}, \cite{ghoshal2021estimating}, \cite{abdar2021uncertainty}, \cite{jungo2020analyzing}, \cite{wang2019aleatoric}, \cite{rkaczkowski2019ara}, \cite{thagaard2020can},  \cite{asgharnezhad2022objective},  \cite{iwamoto2021improving}, \cite{mcclure2019knowing},\cite{molle2019quantifying},\cite{zou2022tbrats}, \cite{mobiny2019risk}, \cite{calderon2021improving-covid}, \cite{calderon2021improving-mammo}, \cite{pocevivciute2022generalisation}, \cite{ahsan2022active},\cite{abdar2021uncertaintyfusenet},\cite{judge2022crisp} \\ \midrule
\textbf{Rejection}           & \cite{nair2020exploring}, \cite{zhang2022epistemic}, \cite{ghoshal2021estimating}, \cite{ozdemir20193d},  \cite{sander2019towards},\cite{abdar2021uncertainty},\cite{tousignant2019prediction}, \cite{filos2019systematic},\cite{abideen2020uncertainty}, \cite{ayhan2018test},\cite{leibig2017leveraging},\cite{mehta2020uncertainty},\cite{mckinley2020uncertainty},\cite{yang2021exploring},\cite{vu2020multi}, \cite{herzog2020integrating},\cite{diao2022unified}, \cite{song2021bayesian},\cite{carneiro2020deep},\cite{mobiny2019risk}, \cite{ghesu2019quantifying},\cite{ghesu2021quantifying}, \cite{combalia2020uncertainty}, \cite{ayhan2020expert},\cite{rosas2021asymmetric}, \cite{rajaraman2022uncertainty}, \cite{tardy2019uncertainty},\cite{yang2021uncertainty},\cite{lambert2022beyond} \\ \midrule
\textbf{OOD detection} &\cite{karimi2020improving},\cite{diao2022unified},\cite{thagaard2020can}, \cite{linmans2020efficient}, \cite{berger2021confidence}, \cite{combalia2020uncertainty}, \cite{tardy2019uncertainty} \\ \midrule
\textbf{Quality Control}  &\cite{ghosal2021uncertainty}, \cite{balagopal2021deep}, \cite{wang2019aleatoric},\cite{mehrtash2020confidence}, \cite{roy2019bayesian}, \cite{devries2018leveraging}, \cite{jungo2018uncertainty},  \cite{jungo2020analyzing}, \cite{hiasa2019automated},\cite{orlando2019u2},  \cite{pan2019prostate}, \cite{mcclure2019knowing}, \cite{hoebel2020exploration},  \cite{shaw2021decoupled}, \cite{rosas2021asymmetric}, \cite{wang2019automatic},\cite{judge2022crisp},\cite{kushibar2022layer} \\ \midrule
\textbf{Label distribution}  & \cite{jungo2018effect}, \cite{baumgartner2019phiseg},  \cite{hu2019supervised}, \cite{li2020uncertainty}, \cite{kohl2018probabilistic}, \cite{kohl2018probabilistic},\cite{gantenbein2020revphiseg},  \cite{monteiro2020stochastic},  \cite{selvan2020uncertainty},  \cite{yang2022uncertainty},\cite{cetindag2022meta}, \cite{bhat2022extending},  \cite{pal2022holistic}, \cite{silva2021using} \\ \bottomrule
\end{longtable}
\end{center}
\end{small}
}

\newpage
%%Harvard
\bibliographystyle{model2-names.bst}
\bibliography{biblio}

\end{document}